\documentclass[aps,pre,twocolumn,superscriptaddress,showkeys, english,showpacs]{revtex4-1}
\usepackage{amsmath}
\usepackage{graphicx}
\usepackage{appendix}
\usepackage{color}
\definecolor{dgreen}{rgb}{0,0.7,0}
\def\redw#1{{\color{red} #1}}

\begin{document}
 \global\long\def\e{\varepsilon}
 \global\long\def\d{\delta}
 \global\long\def\D{\Delta}
  \global\long\def\G{\Gamma}
 \global\long\def\r{\rho}
 \global\long\def\a{\alpha}
  \global\long\def\b{\beta}
 \global\long\def\p{\psi}
 \global\long\def\vf{\varphi}
 \global\long\def\f{\phi}
 \global\long\def\ra{\rightarrow}
 \global\long\def\P{\Psi}
 \global\long\def\m{\mu}
 \global\long\def\n{\nu}
 \global\long\def\g{\gamma}
 \global\long\def\c{\chi}
 \global\long\def\dif{\mathbf{d}}
 \global\long\def\dd{\text{d}}
 \global\long\def\el{\ell}
 \global\long\def\pd#1#2{\frac{\partial#1}{\partial#2}}
 \global\long\def\s{\sigma}
  \global\long\def\S{\Sigma}
 \global\long\def\x{\xi}
 \global\long\def\var{\mbox{var}}
  \global\long\def\cd{\cdot}
   \global\long\def\ord#1{\mathcal{O}\left(#1\right)}
    \global\long\def\asy{\cong}

\title{Derivation of fluctuating hydrodynamics and crossover from diffusive
to anomalous transport in a hard-particle gas}

\author{A. Miron}
\address{Department of Physics of Complex Systems, Weizmann
Institute of Science, Rehovot 76100, Israel}
\author{J. Cividini}
\address{Department of Physics of Complex Systems, Weizmann
Institute of Science, Rehovot 76100, Israel}
\author{Anupam Kundu}
\address{International center for theoretical sciences, TIFR,
Bangalore - 560012, India}
\author{David Mukamel}
\address{Department of Physics of Complex Systems, Weizmann
Institute of Science, Rehovot 76100, Israel}

\begin{abstract}
\noindent 
A recently developed non-linear fluctuating hydrodynamics
theory has been quite successful in describing various features of
anomalous energy transport. However the diffusion and the noise terms
present in this theory are not derived from microscopic descriptions
but rather added phenomenologically. We here derive these hydrodynamic
equations with explicit calculation of the diffusion and noise terms
in a one-dimensional model. We show that in this model the energy
current scales anomalously with system size $L$ as $\sim L^{-2/3}$
in the leading order with a diffusive correction of order $\sim L^{-1}$.
The crossover length $\ell_{c}$ from diffusive to anomalous transport
is expressed in terms of microscopic parameters. Our theoretical predictions
are verified numerically. 
\end{abstract}
\maketitle
\section{Introduction}
\label{introduction}
\noindent 
Often it is observed in many one dimensional systems that
energy transport is not described by Fourier's law, i.e the stationary
current $J_{e}$ does not decay as $J_{e}\propto-\frac{\Delta T}{L}$
for large system size $L$ and small temperature difference $\Delta T$
\cite{NarayanRG,LepriMomExchange,dhar2008heat,Lepri20031,0295-5075-43-3-271,PhysRevLett.92.074302,PhysRevE.68.067102,Lepri2016}.
This phenomenon is manifested by an anomalous asymptotic scaling of
the stationary current $J_{e}\propto-\frac{\Delta T}{L^{1-\alpha}}$
where $0<\alpha\leq1$, power-law decay of the equilibrium current-current
auto-correlations, super-diffusive spreading of local energy perturbations
and non-linear temperature profiles \cite{Lepri2016}.

Recent progress, referred to as non-linear fluctuating hydrodynamics
(NFH) \cite{Spohn2014,van2012exact,Spohn2016,MendlSpohn}, provides
a rather successful theoretical framework for understanding various
aspects of anomalous transport and related phenomena. This theory
describes the dynamics of fluctuations about the equilibrium state
at a nonlinear level, formulated in terms of hydrodynamic (HD) equations
for the conserved fields. In this theory one starts with Euler equations
for the conserved fields into which diffusion and noise terms, satisfying
a fluctuation-dissipation relation (FDR), are added phenomenologically.
While noise and diffusion terms are crucial for deriving the leading
anomalous behavior, remarkably, their explicit values do not affect
the leading anomalous energy current. On the other hand, they do enter
into the next-to-leading contribution which controls the crossover
behavior from finite $L$ to the asymptotic regime. Thus, knowing
the diffusion coefficient is important for reliably analyzing heat
transport data in experiments \cite{Chang,hsiao2013observation} and
in numerical simulations where it is often hard to reach the asymptotic
regime \cite{mai2007t}. It would thus be of great interest to derive
explicit expressions for the diffusion and noise terms in the NFH
equations, starting from a microscopic description.

\begin{figure}[t]
\includegraphics[scale=0.12]{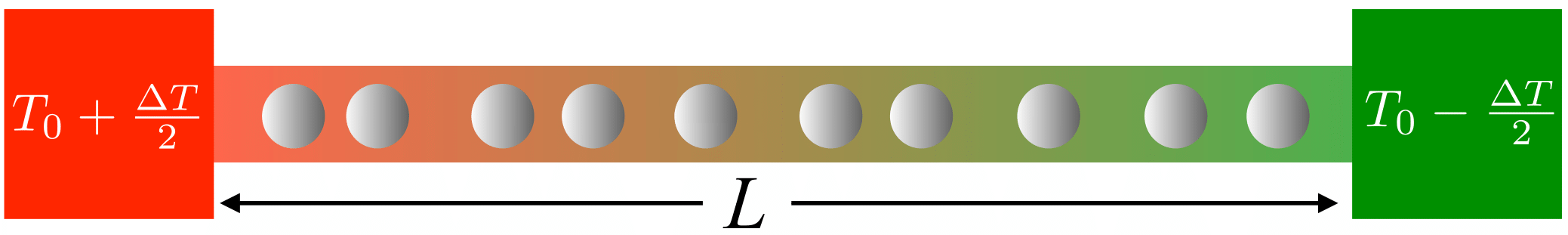} \caption{A gas of unit-mass particles undergoing TPCs, attached to hot and
cold Maxwell heat baths at its two ends.}
\label{fig:model} 
\end{figure}

In this paper, we derive the noise and diffusion terms and study
the crossover behavior in the context of a one-dimensional stochastic
gas model. The model consists of $N$ unit-mass point-particles inside
an interval of size $L$, attached to two Maxwell thermostats \cite{dhar2008heat}
of temperatures $T_{0}\pm\frac{\D T}{2}$ at its ends (see fig. \ref{fig:model}).
The particles undergo stochastic collisions at a constant rate while
evolving ballistically in between collisions. In order to allow for
mixing among the momenta, we consider momentum and energy conserving
collisions involving three neighboring particles. Hereafter, we refer
to this system as the three particle collision (TPC) model. Such three
particle collisions have been considered in several other contexts
\cite{Ma1983,basile2006momentum,iubini2014coarsening,szavits2014constraint}.

\section{Main results}
\label{summary}
Starting from the appropriate master equation, we obtain a ``noisy''
Boltzmann equation, which is then used to derive the NFH equations
for the conserved fields with explicit expressions of the diffusion
and the noise terms. The TPC model has three conserved fields, namely
the particle density $\r(x,t)$, the momentum density $\r u$ and
the energy density $\r e$ where $u(x,t)$ and $e(x,t)$ are the average
momentum and energy per particle, respectively. Applying the NFH framework
to these equations, one finds that the stationary energy current asymptotically
decays as $L^{-2/3}$. However, the fact that the diffusion constant
can be explicitly computed for the TPC model makes it particularly
appealing for studying the diffusive corrections to the leading anomalous
behavior. Thus, accounting for both leading and diffusive contributions
enables one to observe a crossover from one regime to the other upon
varying $L$.

\begin{figure}[t]
\includegraphics[scale=0.55]{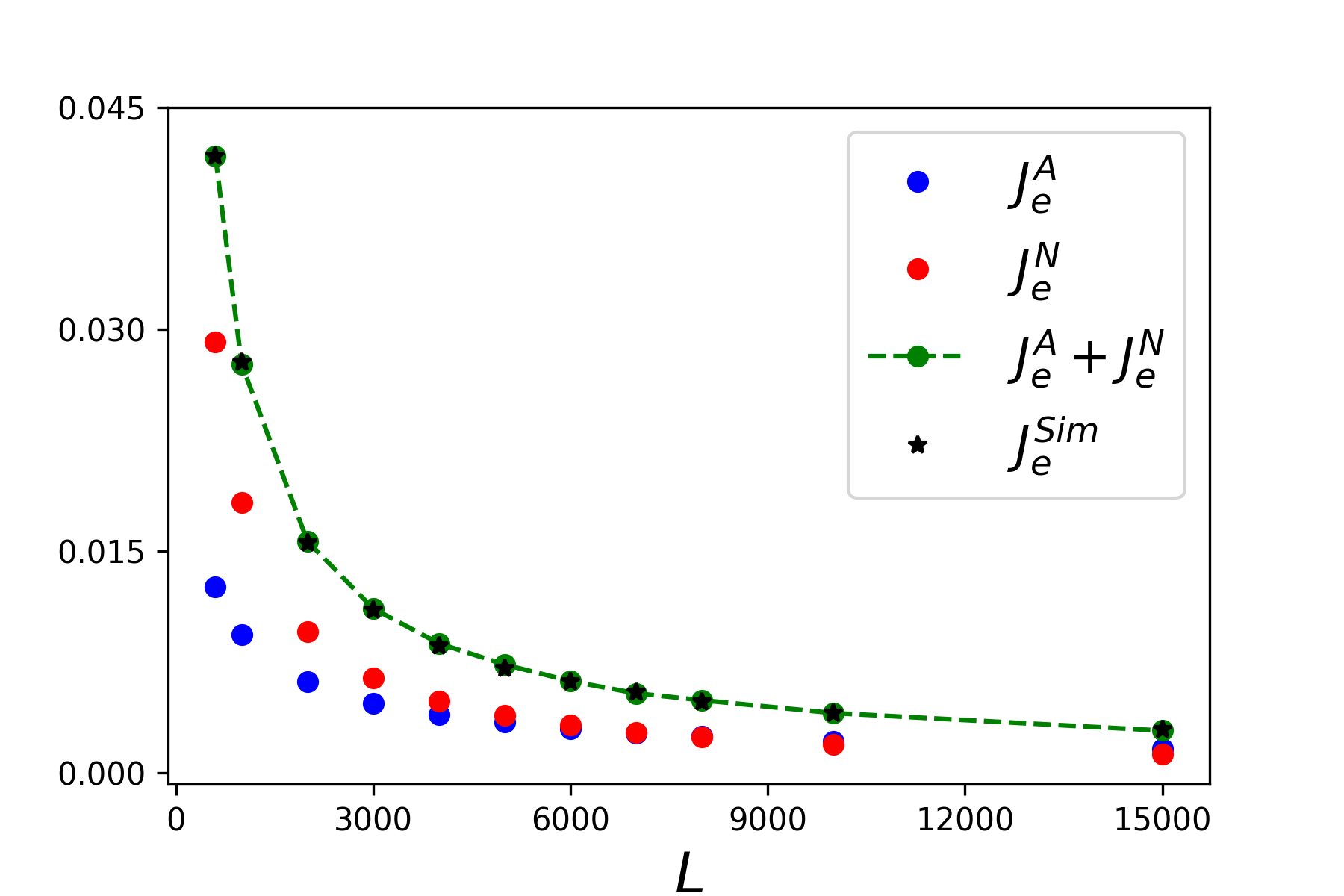} \caption{The stationary current $J_{e}$ as a function of system size $L$
for parameters $\protect\r_{0}=0.5$, $T_{0}=10$ and $\nu_{0}=0.79$.
The black stars ($J_{sim}^{e}$) are direct simulation results, the
blue and red points are the theoretical normal ($J_{e}^{N}\sim L^{-1}$)
and anomalous ($J_{e}^{A}\sim L^{-2/3}$) currents appearing in $\left(1\right)$
and the dashed green line is their sum. We note that the temperature
difference $\protect\D T$ used to compute $J_{e}^{N}$ and $J_{e}^{A}$
is the bulk temperature difference, determined by taking a linear
fit of the bulk temperature profile, extrapolating the linear temperature
profile to the boundaries of the system and computing the difference. }
\label{fig:Je vs L} 
\end{figure}

We find that for small $\Delta T$, the stationary energy current
$J_{e}$ can be written as the sum of a diffusive (normal) part $J_{e}^{N}\sim L^{-1}$
and an anomalous part $J_{e}^{A}\sim L^{-2/3}$ : 
\begin{align}
J_{e}=J_{e}^{N}+J_{e}^{A}=-D\left(1+\left(\frac{L}{\el_{c}}\right)^{1/3}\right)\frac{\D T}{L}
\label{eq:energy_current}
\end{align}
where $D$ is the energy diffusion coefficient and $\el_{c}$ is the
length-scale at which the crossover from diffusive to anomalous transport
takes place. Explicit expressions of $D$ and $\ell_{c}$ are given
in terms of the model parameters in \eqref{eq:Ma's current} and \eqref{eq:crossover_length-1-1}
respectively. It is evident from the expression of $J_{e}$ in \eqref{eq:energy_current}
that for $L\ll\ell_{c}$ the transport is diffusive whereas for $L\gg\ell_{c}$
it is anomalous. In fig. \ref{fig:Je vs L}, $J_{e}$ is plotted against
$L$ for a given set of the model parameters which determine the value
of $D$ and $\el_{c}$, supporting Eq \eqref{eq:energy_current}.
A collapse of the properly scaled $L$ dependence of the current curve
for a large number of sets of model parameters is found in fig. \ref{fig:cross}.
In this figure, it is demonstrated that for different sets of model
parameters the current scaling falls either in the normal regime or
in the anomalous regime. However, finding a single set of model parameters
for which the crossover is evident proved to be numerically difficult.

The paper is organised as follows: 
We proceed by first deriving a stochastic Langevin-Boltzmann (LB) equation in Sec.~\ref{L-B-eq} 
for the empirical density $f(x,p,t)$ which counts the number of particles per unit volume 
of the phase space around the point $(x,p)$. In the next section \ref{SHD}, we make an ansatz 
for the solution of the LB equation, which is then used to derive the stochastic hydrodynamic equations 
for the conserved density, momentum and energy fields.  Our final aim is to compute the system size dependence of the current in NESS with the diffusive correction added to the leading anomalous contribution. This is achieved in Sec.~\ref{LRT} where, starting from the Fokker-Plank equation of the microscopic $N$ particle distribution, we establish a novel linear response theory which expresses the current in NESS as the time integral of the equilibrium current-current correlations  through Green-Kubo formula. These correlations among the currents are, in turn, related to the correlations among the conserved field densities. Finally, these conserved field correlations are evaluated by using the stochastic hydrodynamic equations and extending the mode-coupling procedure to include the desired diffusive correction. As mentioned earlier, this diffusive correction allows us to study a crossover from normal to anomalous transport with increasing system size.  We verify and establish this crossover through extensive numerical simulation. In Sec.~\ref{simulation} we provide the details of our simulation procedure which is followed by our conclusion in Sec.~\ref{conclusion}. 

\section{Derivation of the Langevin-Boltzmann Equation}
\label{L-B-eq}
\noindent
We start by deriving the noisy Langevin-Boltzmann equation for the
empirical density $f(x,p,t)$ at the phase space point $(x,p)$. This
derivation follows the procedure given in \cite{Broeck}.
We first divide the full one particle phase space $(x,p)$ into a large number of
non-overlapping phase-space cells $[x,x+\dd x]\times[p,p+\dd p]$.
As the particles in the TPC gas are evolving with time, the particles
in neighboring phase-space cells get exchanged stochastically and
thus changing the number $N_{(x,p)}(t)=f(x,p,t)\dd x\dd p$ of particles
in these cells.  The state of the system at time $t$ is completely specified by the
set $\left\{ N_{x,p}\left(t\right)\right\} $ and its evolution is
described by a master equation for the joint probability distribution
$P\left(\left\{ N_{x,p}\right\} ;t\right)$ 
\begin{equation}
\partial_{t}P\left(\left\{ N_{x,p}\right\} ;t\right)=\sum_{\left\{ N'_{x,p}\right\} }W_{\left(\left\{ N'_{x,p}\right\} \to\left\{ N_{x,p}\right\} \right)}P\left(\left\{ N'_{x,p}\right\} \right)
\label{eq:mastereq}
\end{equation}
where $W_{\left(\left\{ N'_{x,p}\right\} \to\left\{ N_{x,p}\right\} \right)}$
is the transition rate from the state $\left\{ N'_{x,p}\right\} $
to the state $\left\{ N_{x,p}\right\} $. The transition rate $W_{\left(\left\{ N'_{x,p}\right\} \to\left\{ N_{x,p}\right\} \right)}$
is composed of two contributions: the first describes the drift of
particles along the $x$ axis between adjacent phase-space cells,
i.e $N_{x,p},N_{x+\dif x,p}\ra N_{x,p}-1,N_{x+\dif x,p}+1$, and the
second describes momenta-mixing collisions between triplets of particles
occupying the same $x$ cell, i.e $N_{x,p},N_{x,p'},N_{x,p''},N_{x,q},N_{x,q'},N_{x,q''}\ra N_{x,p}+1,N_{x,p'}+1,N_{x,p''}+1,N_{x,q}-1,N_{x,q'}-1,N_{x,q''}-1$.
One can rewrite the above master equation as 
\begin{align}
\partial_{t}P\left(\left\{ N_{x,p}\right\} ;t\right)=\left(K^{Drift}+K^{Coll}\right)P\left(\left\{ N_{x,p}\right\} ;t\right).\label{eq:mastereq-1}
\end{align}
Explicit expressions of the drift term $K^{Drift}P\left(\left\{ N_{x,p}\right\} \right)$ and the collision term $K^{Coll}P\left(\left\{ N_{x,p}\right\} \right)$ are given in \eqref{eq:drift} and \eqref{eq:collision rate} respectively. 
From this master equation under diffusion approximation
and in the continuum limit, one obtains a Fokker-Plank equation for
the empirical densities $f(x,p,t)$, which corresponds to the following
Langevin-Boltzmann equation (see Appendix-\ref{LB-eq-main} for the derivation)
\begin{align}
\partial_{t}f+p\partial_{x}f=(\partial_{t}f)_{c}+\chi,\label{eq:Boltzmann_eqn}
\end{align}
where $\c\left(x,p,t\right)$ is a zero-mean Gaussian noise satisfying
\begin{align}
\left\langle \c\left(x,p,t\right)\c\left(x',p',t'\right)\right\rangle =\overline{C}\left(p,p'; f \right)\delta(x-x')\delta(t-t'). \nonumber
\end{align}
and is detailed for the present model in Appendix-\ref{FP-for-dist}. 
Here $\overline{C}\left(p,p';f\right)$ is a functional of $f(x,p,t)$
and is provided explicitly in Eq.~\eqref{eq:noisecorr}. 
Note that the noise-free part of \eqref{eq:Boltzmann_eqn} is the
Boltzmann equation which provides the regular evolution of $f(x,p,t)$
whereas the noise term $\chi\left(x,p,t\right)$ describes fluctuations
around this regular evolution. The collision term $(\partial_{t}f)_{c}$
on the right hand side of \eqref{eq:Boltzmann_eqn} describes three
particle collisions occurring at position $x$ and is given by 
\begin{align}
 & (\partial_{t}f)_{c}=\int dp'dp''dqdq'dq''R({\bf p}|{\bf q})(f_{q}f_{q'}f_{q''}-f_{p}f_{p'}f_{p''})\nonumber \\
 & \text{where},~~~R\left(\boldsymbol{p}|\boldsymbol{q}\right)=\g\d\left(P-Q\right)\d\left(E_{\boldsymbol{p}}-E_{\boldsymbol{q}}\right),\label{eq:collision_term}
\end{align}
and $\gamma$ is a constant. Here, $\boldsymbol{p}\equiv\left(p,p',p''\right)$,
$f_{p}\equiv f(x,p,t)$, $P=p+p'+p''$, $Q=q+q'+q''$, $E_{\boldsymbol{p}}=\frac{1}{2}\left(p^{2}+p'^{2}+p''^{2}\right)$
and similarly $E_{\boldsymbol{q}}$. The $\delta$-functions appearing
in the collision kernel $R\left(\boldsymbol{p}|\boldsymbol{q}\right)$
ensure momentum and energy conservation at each collision. At this
point, one needs to solve \eqref{eq:Boltzmann_eqn} for $f(x,p,t)$
along with the noise. However, this task is not straightforward as
Eq. \eqref{eq:Boltzmann_eqn} is nonlinear.

\section{Derivation of the stochastic hydrodynamic equations}
\label{SHD}
\noindent
Let us first consider the significantly simpler linearized, noise-free
version of \eqref{eq:Boltzmann_eqn}. This is done in a non-equilibrium
setting characterized by temperature and density profiles $T\left(x\right)$
and $\r\left(x\right)$ respectively, at constant pressure $P=\r\left(x\right)$T$\left(x\right)$.
Expanding around the local equilibrium (LE) state, Ma \cite{Ma1983}
computed the stationary $f\left(x,p\right)$, to linear order in $\partial_{x}T$,
as 
\begin{align}
f\left(x,p\right)=\frac{\r_{0}e^{-\frac{p^{2}}{2T_{0}}}}{\sqrt{2\pi T_{0}}}\left(1-\frac{9\sqrt{3}}{2\n_{0}\sqrt{2T_{0}}}\vf\left(\frac{p}{\sqrt{T_{0}}}\right)\partial_{x}T\right),\label{eq:eq:Ma's dist}
\end{align}
where $\nu_{0}=\frac{2\pi}{\sqrt{3}}\g\r{}_{0}^{2}$ is the collision
rate, $\rho_{0}=\frac{N}{L}$ is the average density and $\vf\left(p\right)=\frac{p\left(p^{2}-3\right)}{\sqrt{6}}$.
Using the above distribution, one finds a normal current 
\begin{align}
 & J_{e}^{N}=-D\frac{\Delta T}{L},\text{ }\text{where, }D=\frac{27\rho_{0}T_{0}}{4\nu_{0}}.\label{eq:Ma's current}
\end{align}
Note that the energy current \eqref{eq:Ma's current} derived from
the noise-free problem above does not contain the expected anomalous
contribution mentioned in Eq. \eqref{eq:energy_current}. To go beyond
this simple approach, one must study the stochastic evolution of the
conserved fields: $\rho\left(x,t\right)$, $\rho\left(x,t\right)u\left(x,t\right)$
and $\rho\left(x,t\right)e\left(x,t\right)$ which can be obtained
from the Langevin-Boltzmann Eq \eqref{eq:Boltzmann_eqn}. Assuming
that the noisy evolution of the system in the non-stationary regime
can be described as an evolving LE picture at HD length and time scales,
we make the following ansatz for the solution of \eqref{eq:Boltzmann_eqn}
\begin{equation}
f\left(x,p,t\right)=\frac{\r}{\sqrt{2\pi T}}e^{-\frac{\left(p-u\right)^{2}}{2T}}\left(1+\p\text{ }\vf\left(\frac{p-u}{\sqrt{T}}\right)\right)\label{eq:Ansatz}
\end{equation}
where now the fields $\r=\r\left(x,t\right)$, $u=u\left(x,t\right)$,
$T=T\left(x,t\right)$ and $\p=\p\left(x,t\right)$ fluctuate in time
and space due to the noise $\chi$ in Eq \eqref{eq:Boltzmann_eqn}.
These four fields are related to the four (empirical) moments $\mu_{n}\left(x,t\right)=\int\dd pp^{n}f(x,p,t)$
for $n=0,1,2,3$. The evolution equations for $\mu_{n}\left(x,t\right)$
are next derived from Eqs \eqref{eq:Boltzmann_eqn} and \eqref{eq:Ansatz},
yielding the following HD equations for the three conserved fields
$\rho(x,t)$, $\rho(x,t)u(x,t)$, $\rho(x,t)e(x,t)$: 
\begin{align}
 & \partial_{t}\r+\partial_{x}\left(\r u\right)=0\nonumber \\
 & \partial_{t}\left(\r u\right)+2\partial_{x}\left(\r e\right)=0\label{HD-conserved-qtity}\\
 & \partial_{t}\left(\r e\right)+\partial_{x}\left(\sqrt{\frac{3}{2}}\r\p\left(2e-u^{2}\right)^{\frac{3}{2}}+\r u\left(3e-u^{2}\right)\right)=0,\nonumber 
\end{align}
and an equation for the non-conserved field $\psi(x,t)$ 
\begin{align}
\begin{split} & \partial_{t}\r\left(\sqrt{6}\p g_{-}^{\frac{3}{2}}+2u\left(e+g_{-}\right)\right)+\partial_{x}\left(\sqrt{96}\r ug_{-}^{\frac{3}{2}}\p\right)\\
 & =-2\partial_{x}\r\left(g_{+}g_{-}+2e^{2}\right)-\frac{2\sqrt{2}\n_{0}\r g_{-}^{\frac{3}{2}}}{3\sqrt{3}}\p+\sqrt{\sigma}\xi.
\end{split}
\label{eq:psi}
\end{align}
Here $T=2e-u^{2},\text{ }g_{\pm}=2e\pm u^{2}$ and the noise term
is $\xi=\int\dd p\text{ }p^{3}\chi(x,p,t)$ with $\sigma=8\nu_{0}\r_{0}T_{0}^{3}/3$
(see Eq.~\eqref{eq:noise var}).
The equations for the three conserved fields have the expected continuity
form, whereas the equation for $\psi(x,t)$ does not. Also note that
the currents $J_{\rho}=\rho u$ and $J_{\rho u}=\rho e$, associated
with the fields $\rho$ and $\rho u$ respectively, are themselves
conserved. Hence, they do not contain explicit noise terms. On the
other hand, the current in the $\rho e$ equation does contain noise
and dissipation terms through $\psi(x,t)$.

To proceed, we expand the fields in small fluctuations around their
global equilibrium values: $\rho(x,t)\to\rho_{0}+\rho(x,t)$, $u(x,t)\to0+u(x,t)$,
$e(x,t)\to e_{0}+e(x,t)$ and $\psi(x,t)\to0+\psi(x,t)$ (denoting
the fluctuations by the same symbols) and keep only terms of linear
order in fluctuations, obtaining linear fluctuating HD equations.
Since the field $\psi(x,t)$ is not conserved, it evolves on a time
scale of order $\sim\nu_{0}^{-1}$, much shorter than the HD time
scale $\mathcal{O}(L)$ over which the conserved quantities evolve.
This implies $\partial_{t}\psi=\mathcal{O}(\nu_{0}^{-2})$ and so
\begin{equation}
\p\left(x,t\right)=\frac{3\sqrt{3\sigma}}{8\nu_{0}\r_{0}\sqrt{e_{0}^{3}}}\xi-\frac{9\sqrt{3}}{2\nu_{0}\sqrt{e_{0}}}\partial_{x}e+O(\nu_{0}^{-2}),\label{dpsi}
\end{equation}
where $\xi\left(x,t\right)$ is a zero-mean Gaussian white noise with
$\left\langle \xi\left(x,t\right)\xi\left(x',t'\right)\right\rangle =\d\left(x-x'\right)\d\left(t-t'\right)$.
Substituting \eqref{dpsi} into the linearized equation for $e\left(x,t\right)$
yields 
\begin{equation}
\r_{0}\partial_{t}e+\partial_{x}\left(2\r_{0}e_{0}u-2D\partial_{x}e+\sqrt{\Sigma}\xi\right)\simeq0\label{eq:lin_e}
\end{equation}
where $\Sigma=81\sigma/(16\nu_{0}^{2})$. Note that the diffusion
and the noise terms in Eq \eqref{eq:lin_e} satisfy the FDR 
\begin{align}
\frac{\Sigma}{4D}\equiv\text{var}\left[e(x,t)\right]=2e_{0}^{2}.\label{FDR-1}
\end{align}
Once the diffusion and noise terms in the linearized HD equations
are obtained, the NFH equations are constructed by reintroducing the
previously neglected second-order conserved field fluctuations (see
Eq.~(\ref{eq:FHD-real-space}). 

In the NFH theory \cite{Spohn2014,Spohn2016}, the HD equations are
written in the Lagrangian frame in which the conserved quantities
are the stretch field $\ell=\rho^{-1}$, the momentum field $u$ and
the energy field $e$. On the other hand, the HD equations we have
derived in \eqref{HD-conserved-qtity} are expressed in the Eulerian
frame. By making a coordinate transformation from the Eulerian coordinates
$(x,t)$ to the Lagrangian coordinates $(y,t)$ we get (see Appendix-\ref{Eu-to-lag}). 
\begin{align}
 & \partial_{t}\el-\partial_{y}u=0\nonumber \\
 & \partial_{t}u+2\partial_{y}\left(\frac{e}{\ell_{0}}-\frac{e_{0}}{\ell_{0}^{2}}\ell-\frac{e}{\ell_{0}^{2}}\ell-\frac{u^{2}}{2\ell_{0}}\right)=0\label{eq:HD-in-label-space-1}\\
 & \partial_{t}e-\partial_{y}\left(\frac{2D}{\el_{0}^{2}}\partial_{y}e-\frac{2e_{0}u}{\ell_{0}}+\frac{2u}{\el_{0}^{2}}\left(e_{0}\el-e\el_{0}\right)
+\sqrt{\bar{\Sigma}}\bar{\xi}\right)=0\nonumber 
\end{align}
where $\bar{\Sigma}=\r_{0}\Sigma$ and $\bar{\xi}(y,t)=\xi/\sqrt{\rho_{0}}$
has zero mean and variance $\left\langle \bar{\xi}\left(y,t\right)\bar{\xi}\left(y',t'\right)\right\rangle =\d\left(y-y'\right)\d\left(t-t'\right)$.
As before, the diffusion and noise terms are related via FDR in Eq.
\eqref{FDR-1}. Equations \eqref{eq:HD-in-label-space-1} are the
starting-point of the NFH theory \cite{Spohn2014,Spohn2016}. We stress
that, unlike the phenomenological approach taken in the derivation
of the NFH theory, the noise and diffusion terms in \eqref{eq:HD-in-label-space-1}
are derived from a microscopic description of the TPC model. These
equations constitute a significant part of our results.

\noindent 
\begin{figure}[t]
\includegraphics[scale=0.55]{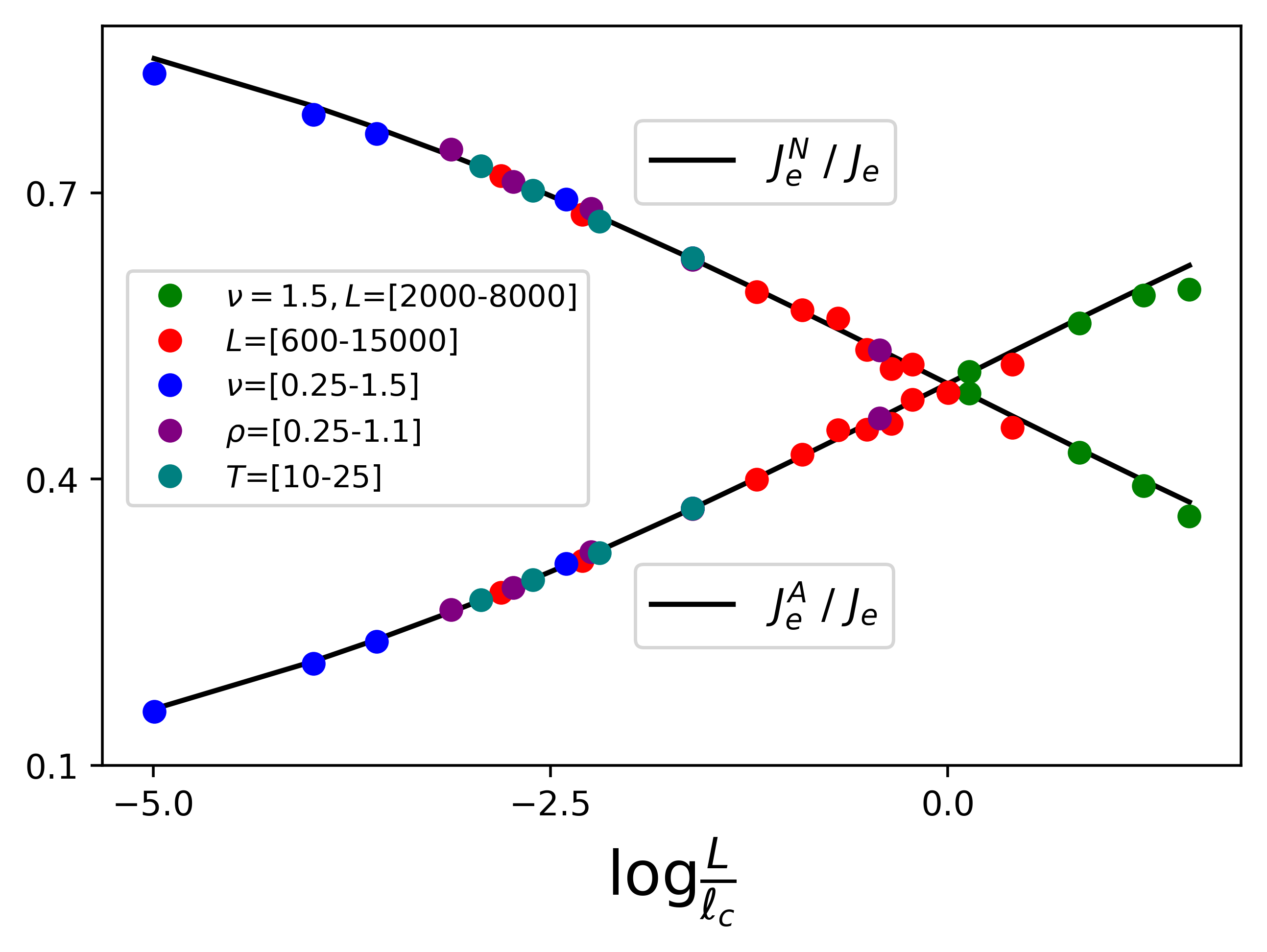} \caption{Data collapse of the ratios $J_{e}^{N}/J_{e}$ and $J_{e}^{A}/J_{e}$
against $u=\log\frac{L}{\protect\el_{c}}$ for various microscopic
parameters and comparison with the collapse functions $J_{e}^{N}/J_{e}=(1+\exp(u/3))$
and $J_{e}^{A}/J_{e}=(1+\exp(-u/3))$. From \eqref{eq:energy_current},
\eqref{eq:Ma's current} and \eqref{eq:crossover_length-1-1} one
finds $J_{e}$ is a function of various parameters $\rho_{0}$, $T_{0}$,
$\nu_{0}$, $\Delta T$ and $L$. Data points (squares and circles)
are simulated by varying one parameter, keeping others fixed. \label{fig:cross} }
\end{figure}

\section{Linear response theory}
\label{LRT}
\noindent
Our final aim is to obtain the $L$ dependence of the stationary energy
current $J_{e}$. We start from the Fokker-Planck equation
describing the evolution of the $N$-particle distribution function $P\left(\G;t\right)$
\begin{equation}
\partial_{t}P\left(\G;t\right)=\L P\left(\G;t\right)\label{eq:FP}
\end{equation}
where the set of particle positions and momenta is denoted by $\G=\left\{ x_{\el},p_{\el}\right\} _{\el=1}^{N}$.
The FP operator $\L$ is defined by its action on a test function
$g\left(\G\right)$ as 
\begin{equation}
\L g\left(\G\right)=\left[-\sum_{i=1}^{N}p_{i}\partial_{x_{i}}+\L_{coll}\right]g\left(\G\right)\label{eq:LP}
\end{equation}
with $\L_{coll}$ denoting the three-particle collision operator which
conserves momentum and energy. 
Here, we would like to emphasise  that the FP equation \eqref{eq:FP} is different from 
the master equation \eqref{eq:mastereq-1}. The Eq.~\eqref{eq:FP} describes the evolution of the joint distribution of 
the particles in phase space whereas the Eq.~\eqref{eq:mastereq-1} describes the evolution of the probability of the occupation in regions of phase space. In this sense the Eq.~\eqref{eq:FP} is a complete microscopic description whereas the Eq.~\eqref{eq:mastereq-1} provides a coarse grained description which is in principle obtained from the former.

The system is driven out of equilibrium
at time $t=0$ by a small temperature difference. The solution of
Eq. \eqref{eq:FP} can be written as 
\begin{equation}
P\left(\G;t\right)=P_{LE}\left(\G\right)+P_{dev}\left(\G;t\right)\label{eq:Dev}
\end{equation}
where $P_{LE}\left(\G\right)$ is the local equilibrium distribution
and $P_{dev}\left(\G;t\right)$ is a deviation from it. The explicit
expression for $P_{LE}\left(\G\right)$ is given by 
\begin{equation}
P_{LE}\left(\G\right)=\prod_{\el=1}^{N}\frac{\r\left(x_{\el}\right)}{N\sqrt{2\pi T\left(x_{\el}\right)}}e^{-\frac{p_{\el}^{2}}{2T\left(x_{\el}\right)}}.\label{eq:LE}
\end{equation}
Substituting Eq. \eqref{eq:Dev} into Eq. \eqref{eq:FP} yields 
\begin{equation}
\partial_{t}P_{dev}\left(\G;t\right)-\L P_{dev}\left(\G;t\right)=\L P_{LE}\left(\G\right),\label{eq:Dev eq}
\end{equation}
whose formal solution is 
\begin{equation}
P_{dev}\left(\G;t\right)=\int_{0}^{t}\dif t'e^{\L\left(t-t'\right)}\L P_{LE}\left(\G\right),\label{eq:Dev sol}
\end{equation}
with an explicit expression of $\L P_{LE}\left(\G\right)$ given by.
\begin{align}
\L P_{LE}\left(\G\right)=-P_{LE}\left(\G\right)&\sum_{i=1}^{N}\left[\frac{\left(p_{i}^{3}-p_{i}T\left(x_{i}\right)\right)}{2T\left(x_{i}\right)^{2}} \right. \nonumber \\
& \left. \times\partial_{x_{i}}T\left(x_{i}\right)+p_{i}\frac{\partial_{x_{i}}\r\left(x_{i}\right)}{\r(x_i)}\right].\label{eq:LP1}
\end{align}
It is often convenient to express the deviation $P_{dev}$ in terms of the currents generated due to the drive. For the TPC gas, one can easily define the instantaneous particle and energy density currents at position $x$ and time $t$ as 
\begin{equation}
\begin{cases}
j_{\r}\left(x,t\right)\equiv\sum_{\el=1}^{N}\d\left(x_{\el}-x\right)p_{\el}\\
j_{e}\left(x,t\right)\equiv\sum_{\el=1}^{N}\d\left(x_{\el}-x\right)\frac{p_{\el}^{3}}{2}
\end{cases}.\label{eq:Js}
\end{equation}
Using \eqref{eq:Js} and the gas equation of state $P=\r\left(x\right)T\left(x\right)$
(with constant pressure $P$) in \eqref{eq:LP1} gives 
\begin{align}
\begin{split}
\L P_{LE}\left(\G\right)=-&P_{LE}\left(\G\right)\int_{0}^{L}\dif x'\frac{\partial_{x'}T\left(x'\right)}{T\left(x'\right)^{2}} 
 \\ 
& \times \left[j_{e}\left(x',t\right)-\frac{3}{2}T\left(x'\right)j_{\r}\left(x',t\right)\right]
\end{split}
\label{eq:LP2}
\end{align}
where the $\G$ dependence of the currents is implicit in Eq. \eqref{eq:LP2}.
Substituting Eq. \eqref{eq:LP2} into Eq. \eqref{eq:Dev sol} yields
the deviation from the local-equilibrium state 
\begin{align}
P_{dev}\left(\G;t\right)=&-\int_{0}^{t}\dif t'e^{\L\left(t-t'\right)}P_{LE}\left(\G\right)\int_{0}^{L}\dif x'\frac{\partial_{x'}T\left(x'\right)}{T\left(x'\right)^{2}} \nonumber \\ 
& \times \left[j_{\r e}\left(x',t'\right)-\frac{3}{2}T\left(x'\right)j_{\r}\left(x',t'\right)\right].
\label{eq:Dev sol-1}
\end{align}
We are now in a position to compute the deviation of the average of any observable from its value in LE. As we are interested in currents, we compute the (non-equilibrium) average particle density current and the energy
density current, $J_{\r}$ and $J_{e}$ respectively,
\begin{align}
\begin{split}
J_{\r}\left(x,t\right)=&\int\dd\G P_{dev}\left(\G,t\right)j_{\r}\left(x,t\right)  \\
J_{e}\left(x,t\right)=&\int\dd\G P_{dev}\left(\G,t\right)j_{e}\left(x,t\right), 
\end{split}
\label{eq:currents}
\end{align}
in the long time limit $t\ra\infty$ gives 
\begin{align}
J_{\r}(x)=&-\int_{0}^{t}\dif t\int_{0}^{L}\dif x'\left[\left\langle j_{\r}\left(x,t\right)j_{e}\left(x',0\right)\right\rangle _{eq} \right. \nonumber \\ 
&\left.-\frac{3}{2}T\left(x'\right)\left\langle j_{\r}\left(x,t\right)j_{\r}\left(x',0\right)\right\rangle _{eq}\right]\frac{\partial_{x'}T\left(x'\right)}{T\left(x'\right)^{2}}\\
J_{e}(x)=&-\int_{0}^{\infty} \dif t\int_{0}^{L}\dif x'\left[\left\langle j_{e}\left(x,t\right)j_{e}\left(x',0\right)\right\rangle _{eq} \right. \nonumber \\ 
&\left. -\frac{3}{2}T\left(x'\right)\left\langle j_{e}\left(x,t\right)j_{\r}\left(x',0\right)\right\rangle _{eq}\right]\frac{\partial_{x'}T\left(x'\right)}{T\left(x'\right)^{2}}
\label{eq:J's fin}
\end{align}
where $\left\langle \cd\right\rangle _{eq}$ denotes an average with
respect to the local-equilibrium distribution \eqref{eq:LE}.

The baths at the boundaries of our TPC gas do not allow for particle current exchange, we do not have any particle current in the steady state. 
Hence $J_\r=0$ and applying this in \eqref{eq:J's fin} yields the relation 
\begin{align}
&\int_{0}^{\infty}\dif t\int_{0}^{L}\dif x'\left\langle j_{\r}\left(x,t\right)j_{e}\left(x',0\right)\right\rangle _{eq}\frac{\partial_{x'}T\left(x'\right)}{T\left(x'\right)^{2}} \nonumber \\ 
&=\frac{3}{2}\int_{0}^{\infty}\dif t\int_{0}^{L}\dif x'\left\langle j_{\r}\left(x,t\right)j_{\r}\left(x',0\right)\right\rangle _{eq}\frac{\partial_{x'}T\left(x'\right)}{T\left(x'\right)}.\label{eq:J_rho}
\end{align}
Simplifying the expression of $J_{e}(x)$ in Eq. \eqref{eq:J's fin}
with help of Eq. \eqref{eq:J_rho} gives 
\begin{align}
&J_{e}(x)=-\int_{0}^{\infty}\dif t\int_{0}^{L}\dif x'\left[\frac{1}{T\left(x'\right)^{2}}\left\langle j_{e}\left(x,t\right)j_{e}\left(x',0\right)\right\rangle _{eq} \right. \nonumber \\ 
&\left.-\frac{9}{4}\left\langle j_{\r}\left(x,t\right)j_{\r}\left(x',0\right)\right\rangle _{eq}\right]\partial_{x'}T\left(x'\right).\label{eq:J e}
\end{align}
In this context, we are only interested in the linear response thus
only the leading contribution in $\partial_{x}T(x)$ is kept and $J_{e}(x)$
becomes 
\begin{align}
&J_{e}(x)=-\frac{1}{T_{0}^{2}}\int_{0}^{\infty}\dif t\int_{0}^{L}\dif x'\left[\left\langle j_{e}\left(x,t\right)j_{e}\left(x',0\right)\right\rangle _{eq} \right. \nonumber \\ 
&~~~~~~~~~~~~~~\left.-\frac{9T_{0}^{2}}{4}\left\langle j_{\r}\left(x,t\right)j_{\r}\left(x',0\right)\right\rangle _{eq}\right]\partial_{x'}T\left(x'\right).\label{eq:J e fin}
\end{align}
To proceed, we relate the correlation functions of the currents to
correlation functions of the fields. We apply the second
moment sum rule for a general conserved field $a\left(x,t\right)$
and a general scaling function $g\left(x/L\right)$ 
\begin{align}
&\int dx~g\left(x/L\right)x^{2}\partial_{t}^{2}\langle a\left(x,t\right)a\left(0,0\right)\rangle_{eq} \nonumber \\ 
&~~~~\approx2\int dx~g\left(x/L\right)\langle j_{a}\left(x,t\right)j_{a}\left(0,0\right)\rangle_{eq}
\end{align}
where the current $j_{a}\left(x,t\right)$ satisfies the continuity
equation $\partial_{t}a\left(x,t\right)=-\partial_{x}j_{a}\left(x,t\right)$.
This equation is valid in the large $L$ limit, as we have neglected
terms smaller than $\mathcal{O}\left(L^{-1}\right)$. Using this relation in
Eq. \eqref{eq:J e fin} we get 
\begin{align}
&J_{e}\left(x\right)=-\frac{1}{T_{0}^{2}}\int_{0}^{\infty}\dif t~\partial_{t}^{2}\int_{0}^{L}\dif x'\frac{\left(x-x'\right)^{2}}{2} 
\partial_{x'}T\left(x'\right) \label{eq:J e fin-1}  \\ 
&~~~\times \left[\left\langle \r e\left(x,t\right)\r e\left(x',0\right)\right\rangle _{eq} \right.\left.-\frac{9T_{0}^{2}}{4}\left\langle \r\left(x,t\right)\r\left(x',0\right)\right\rangle _{eq}\right]. \nonumber
\end{align}
Note that equation \eqref{eq:J e fin-1} is written in real space
where $j_{e}$ is the conserved current of the energy density $\r e$.
We are interested in expressing our results in the language of the
NFH theory, in which the correlation functions are derived in label
space with $y\in\left[0,N\right]$. The transformation from $(x,t)$
to $(y,t)$ is given in Eq. \eqref{eq:label}in which the fields transform
as $\r(x,t)\to\frac{1}{\el(y,t)}$, $u(x,t)\to u(y,t)$ and $e(x,t)\to e(y,t)$.
Hence, Eq. \eqref{eq:J e fin-1} in label space reads 
\begin{align}
&J_{e}(y)=-\frac{\el_{0}^{2}}{T_{0}^{2}}\int_{0}^{\infty}\dif t~\partial_{t}^{2}\int_{0}^{N}\dif y'\frac{(y-y')^{2}}{2} 
\partial_{y'}T\left(y'\right) \label{eq:J e fin-2}  \\ 
&~~~\times \left[\left\langle \frac{e(y,t)}{\el(y,t)}\frac{e(y',0)}{\el(y',0)}\right\rangle _{eq}-\frac{9T_{0}^{2}}{4}\left\langle \frac{1}{\el(y,t)}\frac{1}{\el(y',0)}\right\rangle _{eq}\right].\nonumber
\end{align}
We now expand the fields in fluctuations around their global equilibrium
values: $\el\to\el_{0}+\delta\el$, $u\to0+\delta u$ and $e\to e_{0}+\delta e$.
Keeping terms up to $\ord{\d^{2}}$ in Eq. \eqref{eq:J e fin-2}, one
finds 
\begin{align}
\begin{split}
J_{e}(y)  =-&\int_{0}^{\infty}\dif t~\partial_{t}^{2}\int_{0}^{N}\dif y'\frac{(y-y')^{2}}{2}\partial_{y'}T\left(y'\right)\\
 &\times\left[\frac{1}{T_{0}^{2}}\left\langle \delta e(y,t)\delta e(y',0)\right\rangle _{eq} \right. \\
 & ~~~~~~-\frac{1}{\el_{0}T_{0}}\left\langle \delta\el(y,t)\delta e(y',0)\right\rangle _{eq} \\ 
 & ~~~~~~~~~~\left. -\frac{2}{\el_{0}^{2}}\left\langle \delta\el(y,t)\delta\el(y',0)\right\rangle _{eq}\right].
\end{split}
\label{eq:J label}
\end{align}
To arrive at the above expression we have used $e_{0}=T_{0}/2$
and the following properties $\langle\delta e\rangle_{eq}=\langle\delta\el\rangle_{eq}=0$,
$\langle\delta e(y,t)\delta\el(y,t)\rangle_{eq}=\langle\delta e(y,0)\delta\el(y,0)\rangle_{eq}=\delta(y-y')$,
$\langle\delta e(y,t)\delta\el(y,0)\rangle_{eq}=\langle\delta\el(y,t)\delta e(y,0)\rangle_{eq}$
and the fact that $\langle\delta e(y,t)^{2}\rangle_{eq}$ and $\langle\delta\el(y,t)^{2}\rangle_{eq}$
are independent of $t$.
Next we compute these correlations among the conserved fields
that appear in the above Eq. \eqref{eq:J label} using the evolution equations
for the HD fields given in Eq. \eqref{FHD-label-space}. In order to do so, 
we, at this stage, connect to the theory of NFH \cite{Spohn2014} in which
linearized evolution equations for the conserved fields $\el\left(y,t\right),u\left(y,t\right)$
and $e\left(y,t\right)$ are first decoupled by a transformation to
the eigenbasis. The eigenmodes $\f_{\a}\left(y,t\right)$ ($\a=\pm1,0$)
are linear combinations of the fields where $\f_{\pm1}\left(y,t\right)$
describe two counter-propagating ``sound modes'' whereas $\f_{0}\left(y,t\right)$
describes the non-propagating ``heat mode''. In NFH the coupling
between the sound and heat modes leads to the super-diffusive scaling
of their correlation functions $f_{\a}\left(y,t\right)\equiv\left\langle \f_{\a}\left(y,t\right)\f_{\a}\left(0,0\right)\right\rangle _{eq}$
\cite{Spohn2016,Spohn2014} where only diagonal correlators
(i.e $\left\langle \f_{\a}\f_{\b}\right\rangle $ with $\a=\b$) are
observed to prevail in the long time limit $t\ra\infty$. The evolution
equations of these correlators are solved in the mode-coupling approximation
for asymptotically long time and large distance regime, revealing
their scaling form. In the TPC model, the set of $f_{\a}\left(y,t\right)$
are related to the conserved field correlators by 
\begin{align}
\begin{split}
\left\langle \d\el\left(y,t\right) \d \el\left(y',0\right)\right\rangle _{eq}  =\frac{\el_{0}^{2}}{6}&\left(f_{-1}\left(y-y',t\right)
+4f_{0}\left(y-y',t\right) \right. \\ 
&~~~\left.+f_{+1}\left(y-y',t\right)\right)\\
\left\langle \d\el\left(y,t\right) \d e\left(y',0\right)\right\rangle _{eq}  =\frac{\el_{0}T_{0}}{6}&\left(f_{-1}\left(y-y',t\right)-2f_{0}\left(y-y',t\right)\right. \\ 
&~~~\left. +f_{+1}\left(y-y',t\right)\right)\\
\left\langle  \d e\left(y,t\right) \d e\left(y',0\right)\right\rangle _{eq}  =\frac{T_{0}^{2}}{6}&\left(f_{-1}\left(y-y',t\right)+f_{0}\left(y-y',t\right) \right. \\ 
&~~~\left. +f_{+1}\left(y-y',t\right)\right)
\end{split}
.\label{eq:correlator transformation}
\end{align}
Note that in the above equations, the correlations among the conserved fields are calculated in global equilibrium characterised by $T_0$, $\ell_0$ and zero average momentum density, while in Eq.~\eqref{eq:J label}, these correlations are evaluated at local equilibrium. Since we are interested in leading orders of $\partial_yT(y)$, it is justified to neglect any corrections of order $\partial_yT(y)$ in density correlation that may be present when computed in actual local equilibrium state.

Using the correlation in Eq.~\eqref{eq:correlator transformation}, 
in Eq. \eqref{eq:J label} and simplifying one obtains 
\begin{align}
J_{\r e}(y)=\frac{3}{2}\int_{0}^{\infty}\dif t~\partial_{t}^{2}&\int_{0}^{N}\dif y' \frac{\left(y-y'\right)^{2}}{2} \nonumber \\
& \times f_{0}\left(y-y',t\right)\partial_{y'}T\left(y'\right).\label{eq:linear resp final}
\end{align}
where the heat-mode correlator $f_{0}(y,t)$ is given by its Fourier
transform $f_{0}(y,t)=\int\dd k~e^{2\pi iky}\hat{f}_{0}(k,t)$. The
leading asymptotic scaling form 
\begin{align}
\hat{f}_{0}\left(k,t\right)\approx e^{-\lambda_{h}\left|k\right|^{5/3}t}\label{fhat-spohn}
\end{align}
was obtained in \cite{Spohn2016,Spohn2014} with $\lambda_{h}\approx0.3898\frac{20\pi^{8/3}}{3^{5/3}\gamma[2/3]}c$
where $c=\r_{0}\sqrt{3T_{0}}$ is the sound velocity in the TPC model
and \redw{$\gamma[x]$} denotes the Gamma function. Since the objective in \cite{Spohn2016, Spohn2014}
was to study the leading anomalous behavior, the sub-leading diffusive
contribution to \eqref{fhat-spohn} was not considered. Here, we are
interested in the correction coming from the diffusion term. It is
easy to show that by keeping the diffusive term in the mode-coupling
equation for $\hat{f}_{0}\left(k,t\right)$ in \cite{Spohn2016,Spohn2014},
the asymptotic form of $\hat{f}_{0}(k,t)$ becomes 
\begin{align}
\hat{f}_{0}\left(k,t\right)=e^{-t\left(\lambda_{h}\left|k\right|^{5/3}+\frac{2}{3}D\r_{0}\left(2\pi k\right)^{2}\right)},\label{fhat}
\end{align}
where $D=\frac{27\rho_{0}T_{0}}{4\nu_{0}}$. Inserting this form of
$\hat{f}_{0}(k,t)$ into Eq. \eqref{eq:linear resp final} and performing
the remaining integrals gives 
\begin{align}
J_{e}\left(y\right)=\frac{3}{2}&\int_{0}^{N}\dif y'\frac{\left(y-y'\right)^{2}}{2}\partial_{y'}T\left(y'\right)\int\dd ke^{-2\pi ik(y-y')}\nonumber \\ 
&\int_{0}^{\infty}\dif t~\partial_{t}^{2}e^{-t\left(\lambda_{h}\left|k\right|^{5/3}+\frac{2}{3}D\r_{0}\left(2\pi k\right)^{2}\right)}.\label{eq:bla}
\end{align}
Using the relation $\int\dd k~e^{2\pi ik\left(y-y'\right)}|k|^{5/3}=-\frac{\Gamma\left(\frac{8}{3}\right)}{\left(2\pi\right)^{8/3}}\left|y-y'\right|^{-8/3}$
and simplifying, the stationary current becomes 
\begin{equation}
J_{e}(z)=-\frac{\r_{0}^{1/3}\sqrt{T_{0}}}{\tilde{C}L^{2/3}}\int_{0}^{1}\dif z'\frac{\partial_{z'}T\left(z'\right)}{\left|z-z'\right|^{2/3}}-\frac{D}{L}\partial_{z}T\left(z\right)\label{eq:J_re_final}
\end{equation}
where $z=y/N$, $\tilde{C}^{-1}\approx2.75013$. Since Eq. \eqref{eq:J_re_final}
is an equation for the stationary average energy current $J_{e}(z)$,
which must be independent of $z$, one can verify that there exists
a temperature profile $T\left(z\right)$ such that the right hand
side is also independent of $z$. Using this fact, one may integrate
both sides of \eqref{eq:J_re_final}, replace the temperature profile
by the scaling function $T\left(z\right)=T_{0}+\D T\text{ }h\left(z\right)$
and finally obtain the announced expression for the stationary energy
current (Eq. \eqref{eq:energy_current}) 
\begin{align}
J_{e}=&-\frac{\r_{0}^{1/3}\sqrt{T_{0}}}{C}\frac{\Delta T}{L^{2/3}}-\frac{D\Delta T}{L} \nonumber \\ 
\equiv &-D\left(1+\left(\frac{L}{\el_{c}}\right)^{1/3}\right)\frac{\D T}{L},\label{J-exp-form}
\end{align}
with the crossover length $\ell_{c}$ given by 
\begin{equation}
\ell_{c}=\left(27C\sqrt{T_{0}}/(4\nu_{0})\right)^{3}\rho_{0}^{2}.\label{eq:crossover_length-1-1}
\end{equation}
and the constant $C$ by 
\begin{equation}
\frac{1}{C}=\frac{3}{\tilde{C}}\int_{0}^{1}\dif z'\left[z'^{1/3}+(1-z')^{1/3}\right]\partial_{z'}h\left(z'\right).\label{eq:done}
\end{equation}

From Eqs. \eqref{eq:energy_current}, \eqref{eq:Ma's current} and
\eqref{eq:crossover_length-1-1}, we see that $J_{e}$ depends explicitly
on the system parameters $\rho_{0}$, $T_{0}$, $\nu_{0}$, $\Delta T$
and $L$. In order to verify the theoretical expressions for $J_{e}^{A}$,$J_{e}^{N}$
and $\ell_{c}$ numerically, we plot the ratios $J_{e}^{N}/J_{e}$
and $J_{e}^{A}/J_{e}$ as a function of $u=\log(L/\ell_{c})$ where
$J_{e}^{A}=J_{e}-J_{e}^{N}$ and $J_{e}^{N}$ are obtained from Eqs.
\eqref{eq:energy_current} and \eqref{eq:Ma's current}. It is clear
from \eqref{eq:energy_current}, \eqref{eq:Ma's current} that $J_{e}^{N}/J_{e}=(1+\exp(u/3))$
and $J_{e}^{A}/J_{e}=(1+\exp(-u/3))$. In fig. \ref{fig:cross} we
indeed see that the data for different set of parameters collapse
on these scaling curves. In order to get the best collapse and matching,
we have fitted the free parameter $C$ that appears in \eqref{eq:crossover_length-1-1}.
Our fitted value for $C=0.83$. 

\section{Discussion on the simulation method}
\label{simulation}
\noindent
We now briefly discuss our simulation procedure. In simulations it
is impossible to implement three particle collisions at a point. Instead
we consider collisions between neighboring particle triplets at a
constant rate $\n_{0}$, which at high average density $\r_{0}$ are
in close proximity to each other. Consequently, momentum and energy
are exchanged by particles located at different positions. This procedure
introduces corrections to the diffusion and noise terms appearing
in the HD equations \eqref{HD-conserved-qtity} which, in turn, contributes
to $J_{e}$. The correction to $J_{e}$ due to this exchange, denoted
by $J_{e}^{x}$, is estimated to be $J_{e}^{x}=\frac{2\n_{0}}{3\r_{0}}\left(\frac{\D T}{L}\right)$
(see \cite{kundu2016long}). In order to minimize the contribution
of exchange in the TPC model simulations, we have carefully selected
parameters such that $J_{e}^{x}/J_{e}\sim10^{-2}$, making $J_{e}^{x}$
effectively negligible. For this reason, $J_{e}^{x}$ is absent from
the results shown in fig. \ref{fig:Je vs L} and fig. \ref{fig:cross}. 

\section{Conclusions}
\label{conclusion}
\noindent
In conclusion, we have studied a one-dimensional stochastic gas model whose simple (three particle) collision mechanism conserves momentum and energy, and breaks integrability while still allowing for analytical treatment. Starting from a microscopic description, we have derived a Langevin-Boltzmann equation with a noise term describing our model and used it to derive NFH equations in which both diffusion coefficient and noise amplitude are clearly related to the microscopic model parameters and satisfy FDR. After establishing a novel linear response theory, we compute the current in NESS using the tools of mode-coupling theory. We provide an expression for the stationary energy current of the model which contains the expected leading anomalous contribution but also a normal correction to it. The crossover between normal and anomalous transport involves a typical length-scale $\ell_c$ of which we provide an explicit expression in terms of the microscopic parameter except for a fitting constant.
We verify this crossover through extensive
numerical simulations. In the present study, we consider reservoirs
which prohibit a stationary particle current. Considering reservoirs
which allow both particle and energy transfer could result in two
stationary currents which is an interesting setup to explore. In general,
boundary conditions are observed to have a noticeable effect in systems
featuring anomalous transport \cite{cividini2017temperature,delfini2010nonequilibrium,lepri2011density,LepriMomExchange}.
However, the precise effect of the boundaries in the TPC model is
still unclear. Moreover, it would also be interesting to extend the
present study to other models, for which the diffusion and the noise
terms could be obtained.

\section{Acknowledgement}
\noindent
We thank H. Posch, H. van-Beijern, H. Spohn, A. Dhar and S. N. Majumdar
for useful discussions. AK and JC thank the Weizmann institute of
science for the hospitality received during their visit. AK also acknowledges
support from DST grant under project No. ECR/2017/000634 and the support
form the project 5604-2 of the Indo-French Centre for the Promotion
of Advanced Research (IFCPAR).

\appendix

\begin{widetext}
\section{Derivation of the Langevin-Boltzmann equation \eqref{eq:Boltzmann_eqn}}
\label{LB-eq-main}
\noindent
We start from the master equation \eqref{eq:mastereq-1} 
\begin{equation}
\partial_{t}P\left(\left\{ N_{x,p}\right\} ;t\right)=\left(K^{Drift}+K^{Coll}\right)P\left(\left\{ N_{x,p}\right\} ;t\right),\label{app:eq:mastereq-1}
\end{equation}
the drift term is given by 
\begin{equation}
K^{Drift}P\left(\left\{ N_{x,p}\right\} \right)=\sum_{x',p'}\left|p'\right|\left[E_{x',p'}^{+}E_{x'+\text{sgn}\left[p'\right]\D x,p'}^{-}-1\right]\frac{N_{x',p'}}{\D x}P\left(\left\{ N_{x,p}\right\} \right)\label{eq:drift}
\end{equation}
and the collision term is given by 
\begin{equation}
K^{Coll}P\left(\left\{ N_{x,p}\right\} \right)=\frac{1}{3}\sum_{x}\sum_{\boldsymbol{q},\boldsymbol{p}}\tilde{\gamma}\tilde{R}_{q,q',q''\to p,p',p''}\left[E_{x,q}^{+}E_{x,q'}^{+}E_{x,q''}^{+}E_{x,p}^{-}E_{x,p'}^{-}E_{x,p''}^{-}-1\right]N_{x,q}N_{x,q'}N_{x,q''}P\left(\left\{ N_{x,p}\right\} \right)\label{eq:collision rate}
\end{equation}
where $\tilde{\g}$ has dimension (time)$^{-1}$ and the step operator
$E_{x,p}^{\pm}$ creates/annihilates a particle at the box labeled
$\left(x,p\right)$, i.e 
\begin{equation}
E_{x,p}^{\pm1}P\left(...,N_{x,p},...\right)=P\left(...,N_{x,p}\pm1,...\right).\label{eq:step}
\end{equation}
The notation $\sum_{\boldsymbol{q},\boldsymbol{p}}$ denotes the sums
over the vectors ${\bf q}=(q,q',q'')$ and ${\bf p}=(p,p',p'')$ where
the momentum components lie $\in\left(-\infty,\infty\right)$ and
the kernel $\tilde{R}$ is given by 
\begin{equation}
\tilde{R}_{q,q',q''\to p,p',p''}=\tilde{R}_{{\bf q}\to{\bf p}}=\delta_{p+p'+p''-q-q'-q''}\delta_{\frac{p^{2}+p'^{2}+p''^{2}}{2}-\frac{q^{2}+q'^{2}+q''^{2}}{2}}.\label{eq:collision}
\end{equation}
Here the $\delta$'s are Kronecker deltas. The 1/3 factor in the definition
of $\overline{\g}$ is taken so that the resulting collision rate
of the collision term in the Langevin-Boltzmann equation $\left(19\right)$
coincides with the collision rate $\g$, as defined in equation $\left(4\right)$.

Following \cite{Broeck}, we next simplify \eqref{eq:mastereq} by
taking the continuum limit. We consider the regime where $\D x\D p$
is large enough such that there are many particles in each phase-space
cell while the spatial size $\D x$ of each cell is much smaller than
the system size $L$. Accordingly, we define the phase-space density
$f_{x,p}\left(t\right)\equiv\frac{N_{x,p}\left(t\right)}{\D x\D p}$
and formulate an evolution equation for it. In this regime the step
operators $E_{x,p}^{\pm}$ can be expressed by 
\begin{equation}
E_{x,p}^{\pm}=\exp\left[\pm\pd{}{N_{x,p}\left(t\right)}\right]=\exp\left[\pm\frac{1}{\D x\D p}\pd{}{f_{x,p}\left(t\right)}\right].\label{E-step}
\end{equation}

The continuum limit is obtained by taking the cell size $\Delta x$
and $\D p$ to zero. In order to get the desired limit, we use the
following prescriptions: 
\begin{align}
\begin{split} & \Delta x\sum_{x}\to\int dx,~~\Delta p\sum_{p}\to\int dp\\
 & \delta_{x,y}\to\Delta x~\delta(x-y),~~\delta_{p,q}\to\Delta p~\delta(p-q)\\
 & f_{x,p}\to f(x,p),~~\pd{}{f_{x,p}}\to\Delta x\Delta p~\frac{\delta}{\delta f(x,p)},\\
 & \text{and},~~\tilde{R}_{{\bf q}\to{\bf p}}\to\Delta p^{3}~\bar{R}({\bf q}\to{\bf p}),
\end{split}
\end{align}
where $f(x,p)$ is now the density function of continuous variables
$x,p$. Similarly, $\bar{R}({\bf q}\to{\bf p})$ is now the collision
kernel of the continuous momenta ${\bf q}$ and ${\bf p}$. In the
continuum limit, the derivative $\pd{}{f_{x,p}}$ becomes a functional
derivative $\frac{\delta}{\delta f(x,p)}$ with respect to the function
$f(x,p)$.

\subsection{Continuum limit of the drift term in \eqref{eq:mastereq-1}}

From \eqref{eq:drift} and \eqref{E-step}, we have 
\begin{align}
K^{Drift} & P\left(\left\{ N_{x,p}\right\} \right)\\
 & =\sum_{x',p'}\left|p'\right|\Delta p~f_{x',p'}\left[\exp\left(\frac{1}{\Delta x\Delta p}\sum_{y}(\delta_{y,x'}-\delta_{y,x'+sgn[p']\Delta x'})\frac{\partial}{\partial f_{y,p'}}\right)-1\right]P\left(\left\{ f_{x,p}\right\} ;t\right)\nonumber \\
 & =\sum_{x',p'}\left|p'\right|\Delta p~f_{x',p'}\left[\exp\left(\int dy(\delta(y-x')-\delta(y-x'-sgn[p']\Delta x'))\frac{\delta}{\delta f(y,p')}\right)-1\right]P\left[f(x,p);t\right]\\
 & =\int dp'~p'~\int dy\int dx'\left(\frac{\partial}{\partial y}\delta(x'-y)\right)f(x',p')\frac{\delta P\left[f(x,p),t\right]}{\delta f(y,p')}+O(\Delta x)\\
 & =-\int dx\int dp\frac{\delta}{\delta f(x,p)}\left(A_{1}^{(1)}(x,p)~P[f(x,p);t]\right),~~\text{where}~~A_{1}^{(1)}(x,p)=-p\frac{\partial}{\partial x}f(x,p).\label{drift-1}
\end{align}

\subsection{Continuum limit of the collision term in \eqref{eq:mastereq-1}}

We now derive the continuum limit of the collision term. Starting
from \eqref{eq:drift} and using the definition of the step operators
in \eqref{E-step}, the collision rate is expanded to second order
is $\left(\D x\D p\right)^{-1}$ 
\begin{align}
K^{Coll}P & =\frac{1}{3}\sum_{x}\sum_{{\bf q},{\bf p}}\tilde{\g}\tilde{R}_{{\bf q}\to{\bf p}}\D x^{3}\D p^{3}f_{x,q}f_{x,q'}f_{x,q''}\nonumber \\
 & ~~~~~~~~~~~~~\times\left[\exp\left(\frac{1}{\D x\D p}\sum_{r}\left(\delta_{r,q}+\delta_{r,q'}+\delta_{r,q''}-\delta_{r,p}-\delta_{r,p'}-\delta_{r,p''}\right)\frac{\partial}{\partial f_{x,r}}\right)-1\right]P(\{f_{x,p}\};t).\\
 & =\frac{1}{3}(\tilde{\g}\D x^{2})\int dx\int d{\bf q}\int d{\bf p}~\bar{R}({\bf q}\to{\bf p})~f(x,q)f(x,q')f(x,q'')\left[\exp\left(\int dr~\D(r,{\bf q,p})\frac{\delta}{\delta f(x,r)}\right)-1\right]P[f(x,p);t],
\nonumber 
\end{align}
where 
\begin{align}
\D(r,{\bf q,p})=\delta(r-q)+\delta(r-q')+\delta(r-q'')-\delta(r-p)-\delta(r-p')-\delta(r-p'').
\end{align}
In the $\tilde{\g}\to\infty$ and $\Delta x^{2}\to0$ limit we keep
$\tilde{\g}\Delta x^{2}=\g$ finite. Note that $\g$ has the dimension
(time $\times$ density$^{2}$)$^{-1}$, as in the Ma's paper \cite{Ma1983}.
Following Brenig \emph{et al.} \cite{Broeck}, we make the diffusion
approximation which amounts to expanding the exponential term in upto
second order. Finally we get 
\begin{align}
K^{Coll}P & =\frac{1}{3}\int dx\int d{\bf q}\int d{\bf p}~\g~\bar{R}({\bf q}\to{\bf p})~f(x,q)f(x,q')f(x,q'')\left[\int dr\D(r,{\bf q,p})\frac{\delta}{\delta f(x,r)}\right.\nonumber \\
 & ~~~~~~~~~~~~~~~~~~~~~~~\left.+\frac{1}{2}\int dr\int dr'\D(r,{\bf q,p})\D(r',{\bf q,p})\frac{\delta^{2}}{\delta f(x,r)\delta f(x,r')}\right]P[f(x,p);t],\nonumber \\
 & =-\int dx\int dp\frac{\delta}{\delta f(x,p)}\left(A_{1}^{(2)}(x,p)P[f(x,p);t]\right)\nonumber \\
 & ~~~~~~~~~~~~~+\frac{1}{2}\int dx\int dx'\int dr\int dr'\frac{\delta^{2}}{\delta f(x,r)\delta f(x,r')}\left(A^{(2)}(x,r,x',r')P[f(x,p);t]\right),
\label{term-2}
\end{align}
where 
\begin{align}
A_{1}^{(2)}(x,p) & =\int dp'\int dp''\int dq\int dq'\int dq''\g~\bar{R}({\bf q\to p})~[f(x,q)f(x,q')f(x,q'')-f(x,p)f(x,p')f(x,p'')],\nonumber \\
A^{(2)}(x,r,x',r') & =\frac{\delta(x-x')}{3}\int dp'\int dp''\int dq\int dq'\int dq''~\g~\bar{R}({\bf q\to p})f(x,q)f(x,q')f(x,q'')\nonumber \\
 & ~~~~~~\times\left[\delta\left(q-r\right)+\delta\left(q'-r\right)+\delta\left(q''-r\right)-\delta\left(p-r\right)-\delta\left(p'-r\right)-\delta\left(p''-r\right)\right]\nonumber \\
 & ~~~~~~\times\left[\delta\left(q-r'\right)+\delta\left(q'-r'\right)+\delta\left(q''-r'\right)-\delta\left(p-r'\right)-\delta\left(p'-r'\right)-\delta\left(p''-r'\right)\right].\label{A-2}
\end{align}

\subsection{Fokker-Plank equation for the density function $f(x,p)$}
\label{FP-for-dist}

\noindent Combining the terms in equations \eqref{drift-1} and \eqref{term-2},
we get the following Fokker-Planck equation as the continuum limit
of the master equation \eqref{eq:mastereq-1} 
\begin{align}
\partial_{t}P\left[\left\{ f\left(x,p\right)\right\} ,t\right] & =-\int\dd x\int\dd p\frac{\delta}{\delta f\left(x,p\right)}\left(A_{1}(x,p)P[f(x,p);t]\right)\nonumber \\
 & ~~~~~~~~~~~~+\frac{1}{2}\int\dd x\int\dd p\int\dif x'\int\dd p'\frac{\delta^{2}}{\delta f\left(x,p\right)\delta f\left(x',p'\right)}\left(A_{2}(x,p,x',p')P[f(x,p);t]\right)\label{eq:FP eqn}
\end{align}
where $A_{1}(x,p)=A_{1}^{(1)}(x,p)+A_{1}^{(2)}(x,p)$ and given explicitly
by 
\begin{align}
A_{1}(x,p)=-p\frac{\partial}{\partial x}f(x,p)+\int dp'\int dp''\int dq\int dq'\int dq''~R({\bf q\to p})~[f(x,q)f(x,q')f(x,q'')-f(x,p)f(x,p')f(x,p'')].\label{eq:A1}
\end{align}
Here $R({\bf q\to p})\equiv R({\bf q|p})=\g\bar{R}({\bf q\to p})$
and $A_{2}(x,p,x',p')$ is given in \eqref{A-2}. Note that the second
term on the right hand side of \eqref{eq:A1} is equal to $\left(\partial_{t}f\right)_{c}$
in equation \eqref{eq:collision_term}. The Langevin-Boltzmann
equation corresponding to the Fokker Planck equation \eqref{eq:FP eqn}
can be identified and it is given by 
\begin{equation}
\partial_{t}f\left(x,p,t\right)=-p\partial_{x}f\left(x,p,t\right)+\left(\partial_{t}f\right)_{c}\left(x,p,t\right)+\c\left(x,p,t\right),\label{eq:LB eqn}
\end{equation}
where $\c\left(x,p,t\right)$ is a mean zero Gaussian white noise
whose properties are determined by $A_{2}(x,p,x',p')$: 
\begin{eqnarray}
\langle\c\left(x,p,t\right)\c\left(x',p',t'\right)\rangle & = & \delta\left(t-t'\right)\delta\left(x-x'\right)A_{2}(x,p,x',p')\nonumber \\
 & = & \frac{1}{3}\delta\left(t-t'\right)\delta\left(x-x'\right)\int\dif p_{1}\dif p_{2}\dif p_{3}\dif q_{1}\dif q_{2}\dif q_{3}R\left(q_{1},q_{2},q_{3}\to p_{1},p_{2},p_{3}\right)f\left(x',q_{1}\right)f\left(x',q_{2}\right)f\left(x',q_{3}\right)\nonumber \\
 &  & \times\left[\delta\left(q_{1}-p\right)+\delta\left(q_{2}-p\right)+\delta\left(q_{3}-p\right)-\delta\left(p_{1}-p\right)-\delta\left(p_{2}-p\right)-\delta\left(p_{3}-p\right)\right]\nonumber \\
 &  & \times\left[\delta\left(q_{1}-p'\right)+\delta\left(q_{2}-p'\right)+\delta\left(q_{3}-p'\right)-\delta\left(p_{1}-p'\right)-\delta\left(p_{2}-p'\right)-\delta\left(p_{3}-p'\right)\right].
 \label{eq:noisecorr}
\end{eqnarray}

In the main text we have used this Langevin-Boltzmann equation \eqref{eq:LB eqn}
to derive equations for the first four moments (Eqs. \eqref{HD-conserved-qtity} and \eqref{eq:psi} of the main text), with the noise $\c$ appearing
only in last equation as $\sqrt{\s}\x\left(x,t\right)=\int\dif pp^{3}\c\left(x,p,t\right)$.
These equations are linearized by expanding the fields in small fluctuations
around their global equilibrium values. In the $\nu_{0}\gg1$ limit,
we derive Eq.~\eqref{dpsi} of the main text for $\p\left(x,t\right)$.
To compute the leading approximation of the noise amplitude $\s\left\langle \x\left(x,t\right)\x\left(x',t\right)\right\rangle =\delta(t-t)\int\dif pp^{3}\int\dif p'p'^{3}A_{2}(x,p,x',p')$
in this limit, we replace $f\left(x,p,t\right)$ appearing in $A_{2}(x,p,x',p')$
by the global equilibrium one-particle distribution $f_{\mathrm{0}}\left(p\right)=\frac{\rho_{0}}{\sqrt{2\pi T_{0}}}e^{-\frac{p^{2}}{2T_{0}}}$.
We denote the corresponding variance by $\s\left\langle \x\left(x,t\right)\x\left(x',t\right)\right\rangle _{0}$
which reads 
\begin{eqnarray}
\s\langle\x\left(x,t\right)\x\left(x',t'\right)\rangle_{\mathrm{0}} & = & \delta\left(t-t'\right)\delta\left(x-x'\right)\frac{2\rho_{0}^{3}T_{0}^{3}\g}{\left(2\pi\right)^{3/2}}\int\dif u\dif u'\dif u''\dif v\dif v'\dif v''e^{-\frac{v^{2}+v'^{2}+v''^{2}}{2}}\nonumber \\
 &  & \times\delta(u+u'+u''-v-v'-v'')\delta\left(\frac{1}{2}(u^{2}+u'^{2}+u''^{2}-v^{2}-v'^{2}-v''^{2})\right)\left[u^{6}+2u^{3}u'^{3}-3u^{3}v^{3}\right]\nonumber \\
 & = & \frac{2\rho{}_{0}^{3}T{}_{0}^{3}\g}{3\left(2\pi\right)^{3/2}}\delta\left(t-t'\right)\delta\left(x-x'\right)\int_{P=-\infty}^{\infty}\dd P\int_{r=0}^{\infty}r\dd re^{-\frac{P^{2}}{6}-\frac{r^{2}}{2}}\frac{\pi^{2}r^{6}}{9}\nonumber \\
 & = & \frac{16\pi\g\rho{}_{0}^{3}T{}_{0}^{3}}{3\sqrt{3}}\delta\left(t-t'\right)\delta\left(x-x'\right)=\frac{8\nu_{0}\rho_{0}T_{0}^{3}}{3}\delta(t-t)\delta(x-x'),\label{eq:noise var}
\end{eqnarray}
where we have used the variables defined in \cite{Ma1983} to carry
out integration. From Eq. \eqref{eq:noise var} we identify the noise
amplitude $\s\equiv\frac{8}{3}\nu_{0}\rho_{0}T_{0}^{3}$ appearing
in equation \eqref{eq:psi} in the main text.

From the linearized equation for the energy density $\r e$ in which
the noise amplitude becomes $\S=\left(\frac{9}{4\n_{0}}\right)^{2}\s$
(see Eq. \eqref{eq:lin_e} of the main text), one can easily verify
that the fluctuation dissipation relation is indeed satisfied. This
is done by computing the variance of the energy $e=\frac{p^{2}}{2}$
using the equilibrium distribution $f_{\mathrm{0}}\left(p\right)$
which yields $\var\left[e\right]=2e^{2}$. Since one independently
has $\frac{\S}{4D}=2e_{0}^{2}$, the expected result is obtained 
\begin{equation}
\frac{\S}{4D}=\var\left[e_{0}\right]=2e_{0}^{2}.\label{eq:sigmaD}
\end{equation}

\section{Transformation Between Eulerian (real space) to Lagrangian (label
space) Coordinates}
\label{Eu-to-lag}

\noindent To easily apply the results of the NFH theory \cite{Spohn2014},
we change our reference frame from the ``real-space'' coordinates
$(x,t)$, where $x\in\left[0,L\right]$, to the ``label-space''
coordinates $(y,t)$ in which $y\in\left[0,N\right]$ is the continuous
particle label. The transformation between frames is given explicitly
in Appendix 5 of \cite{Spohn2014} as 
\begin{align}
\int^{x}dx'\r(x',t)=y(x,t),~~\int^{y}dy'\el(y',t)=x(y,t),\label{eq:label}
\end{align}
which is equivalent to 
\begin{equation}
\begin{cases}
\partial_{x}\ra\frac{1}{\el}\partial_{y}\\
\partial_{t}\ra\partial_{t}-\frac{u}{\el}\partial_{y}
\end{cases},\label{eq:label-1}
\end{equation}
where $\r_{0}=N/L$ and the HD fields transform as 
\begin{align}
\begin{cases}
\r(x,t)\to\frac{1}{\el(y,t)}\\
u(x,t)\to u(y,t)\\
e(x,t)\to e(y,t)
\end{cases} & .\label{eq:trans}
\end{align}
Applying this transformation to the real-space HD equations (Eq.~\eqref{HD-conserved-qtity} in the main text with $\p$ replaced by its linearized
form $\left(10\right)$) 
\begin{align}
 & \partial_{t}\r+\partial_{x}\left(\r u\right)=0\nonumber \\
 & \partial_{t}\left(\r u\right)+2\partial_{x}\left(\r e\right)=0\label{eq:FHD-real-space}\\
 & \partial_{t}\left(\r e\right)+\partial_{x}\left(\r u\left(3e-u^{2}\right)-2D\partial_{x}e+\sqrt{\S}\xi\right)=0,\nonumber 
\end{align}
we find 
\begin{align}
 & \partial_{t}\el-\partial_{y}u=0\nonumber \\
 & \partial_{t}u+\partial_{y}\left(\frac{2e-u^{2}}{\el}\right)=0\label{FHD-label-space}\\
 & \partial_{t}e+\partial_{y}\left(\frac{u}{\el}\left(2e-u^{2}\right)-\frac{2D}{\el^{2}}\partial_{y}e+\sqrt{\overline{\S}}\overline{\xi}\right)=0,\nonumber 
\end{align}
where $\overline{\S}=\left(\frac{9}{4\n_{0}}\right)^{2}\r_{0}\s$
and the Gaussian white noise $\bar{\xi}(y,t)$ satisfies 
\begin{equation}
\begin{cases}
\langle\bar{\xi}(y,t)\rangle=0\\
\langle\bar{\xi}(y_{1},t_{1})\bar{\xi}(y_{2},t_{2})\rangle=\delta(y_{1}-y_{2})\delta(t_{1}-t_{2})
\end{cases}.\label{eq:label noise}
\end{equation}
Expanding the currents in Eqs. \eqref{FHD-label-space} to second
order in fluctuations of the conserved fields, we obtain Eqs. \eqref{eq:HD-in-label-space-1}
of the main text.

\end{widetext}

\end{document}